\documentclass{article}
\usepackage{fullpage,natbib,times,epsfig}

\bibpunct[, ]{(}{)}{;}{a}{}{,}

\begin{document}
\title{Peering through the OH Forest: public release of sky-residual subtracted spectra
  for SDSS DR7}
\author{Vivienne Wild$^{1,2}$\thanks{vw@roe.ac.uk},  Paul C. Hewett$^3$ \\
\vspace*{6pt}\\
$^1$Institut d'astrophysique de Paris, 98bis Boulevard Arago, 75014 Paris, France \\
$^2$Institute for Astronomy, Blackford Hill, Edinburgh, EH9 3HJ, UK  \\
$^3$Institute of Astronomy, Madingley Road, Cambridge, CB3 0HA, UK 
}

\maketitle

\begin{abstract}
  
  The Sloan Digital Sky Survey (SDSS) automated spectroscopic
  reduction pipeline provides $>$1.5 million intermediate resolution,
  $R \simeq 2000$, moderate signal-to-noise ratio (SNR), SNR$\sim 15$,
  astronomical spectra of unprecedented homogeneity that cover the
  wavelength range $3800-9200\,$\AA.  However, there remain
  significant systematic residuals in many spectra due to the
  sub-optimal subtraction of the strong OH sky emission lines longward
  of $6700\,$\AA.  The OH sky lines extend over almost half the
  wavelength range of the SDSS spectra, and the SNR over substantial
  wavelength regions in many spectra is reduced by more than a factor
  two over that expected from photon counting statistics. Following
  the OH line subtraction procedure presented in
  \citet[][WH05]{2005MNRAS.358.1083W}, we make available to the
  community sky-residual subtracted spectra for the Sloan Digital Sky
  Survey Data Relase 7. Here we summarise briefly the method,
  including minor changes in the implementation of the procedure with
  respect to WH05. The spectra are suitable for many science
  applications but we highlight some limitations for certain
  investigations. Details of the data model for the sky-residual
  subtracted spectra and instructions on how to access the spectra are
  provided.
\end{abstract}

\section{Introduction}
The Sloan Digital Sky Survey \citep[SDSS;][]{2000AJ....120.1579Y}
represents the most impressive application yet of wide-field
multi-object spectrographs to provide spectroscopic samples of
quasars, galaxies and stars of unprecedented size. The final data
release \citep[DR7;][]{Abazajian:2009p5430} provides $>$1.5 million
science-target spectra, nearly 1.3 million of which form part of the
SDSS Legacy Survey.

Despite the high quality of the SDSS spectra, visual inspection reveals
significant systematic sky-subtraction residuals longward of $6700\,$\AA \ in
many spectra.  For fainter objects in particular, the sky-subtraction residuals,
which exhibit strong non-Gaussian properties, are the dominant source of
uncertainty over a wavelength interval of some $2000\,$\AA.  The wavelength
range affected includes features of significant astrophysical interest; examples
include the Calcium triplet (8500, 8544, $8665\,$\AA), a powerful diagnostic of
stellar populations in low-redshift galaxies, and the H$\beta$ + [OIII] 4960,
$5008\,$\AA \ emission region in active galactic nuclei (AGN) in the redshift
interval $0.4 < z < 0.8$.

The fundamental problem is the difficulty associated with the removal of OH
emission features from spectra in which the line profiles are barely resolved.
Combined with sub-pixel changes in the pixel-to-wavelength calibration between
spectra, sharp (positive/negative) residuals often remain following sky
subtraction.  The sky-subtraction residuals arise from the subtraction of two
essentially identical tooth-comb signatures that have been very slightly
misaligned, leading to well-defined residual patterns.  In WH05 we presented a
method to remove the dominant OH sky-subtraction residuals in the SDSS spectra
based on Principal Component Analysis (PCA).  PCA takes advantage of the
correlations present as a function of wavelength, offering significant
advantages over simply masking the affected pixels.

In this document we describe the release to the astronomical community
of FITS-files containing the sky-residual subtracted fluxes. We
briefly summarise the main steps of the procedure in Section
\ref{sec:method}, including details of the object-feature masks used
and minor alterations to the original method. In Section
\ref{sec:model} we describe the data model and how to access the data.
The core code, written in the IDL language, is also released for
reference.

\section{Method}\label{sec:method}
Here we outline the main points of the procedure, and refer the reader to WH05
for full details.  For each observation of a SDSS spectroscopic plate, 32
fibres, 16 for each of the 2 spectrographs, are normally assigned to blank sky
regions, selected from areas containing no detected objects in the SDSS imaging
survey.  During the SDSS pipeline processing, the sky spectra are combined to
create a ``master-sky'' spectrum for each spectroscopic plate, which is then
scaled and subtracted from each of the 640 spectra, producing 608 sky-subtracted
object spectra and 32 sky-subtracted sky spectra, all of which are part of the
standard SDSS data releases.  Throughout the remainder of the paper we will
refer to the sky-subtracted sky spectra as ``sky spectra'', and the
sky-subtracted object spectra as ``object spectra''.  The purpose of our
technique is to remove the tooth-comb residuals which remain after the SDSS
pipeline master sky subtraction.

The analysis that follows is applied only to the red part of the SDSS spectra
($6700-9180\,$\AA), as the strong emission features blueward of $6700\,$\AA \
are small in number and easily masked.  Additionally, only pixels which may have
been contaminated by night sky lines, termed ``sky-pixels'', are cleaned.  Those
pixels which lie between sky lines, termed ``non-sky pixels'', are not altered
by the procedure and are used simply as reference points during the
reconstruction of the OH features in the sky-pixels.

\subsection{The eigensystem}

The SDSS sky spectra are used to calculate the eigensystem.  The first stage is
to form an estimate of the Poisson noise spectrum for each plate from the median
of the error arrays of the 32 sky spectra on that plate.  The reason for such an
approach, rather than using the individual noise arrays of each sky spectrum, is
the difficulty of separating the sky noise component from the object noise
component in the object error arrays.  Thus, a single Poisson noise spectrum is
used for all objects on the same plate.  This Poisson noise spectrum is
corrected for the rescaling applied by the SDSS reduction pipeline at the
position of the OH lines (see WH05).  With a robust estimate of the Poisson
noise spectrum available for each plate, each sky spectrum is then normalised by
the expected Poisson noise.  Failure to do so would result in over-subtraction
of flux in sky pixels, with ``corrected'' pixels apparently exhibiting
fluctuations below the Poisson limit (see WH05).

The sky-subtraction method then proceeds by performing a PCA of the
Poisson error normalised sky spectra to produce a set of orthogonal
components that provides a compact representation of the systematic
residuals caused by the OH emission lines. In our updated method, a
robust and iterative PCA procedure is used, which is presented in
\citet{Budavari:2009p4530}. The algorithm essentially replaces the
least-squares minimisation solved by traditional PCA with the
minimisation of a new robust function of the data, which employs a
robust Cauchy-type function to limit the impact of outliers. The
iterative nature of the algorithm allows us to greatly reduce the
computational time for the eigensystem; only $\sim3000$
sky spectra are required for the eigensystem to converge.

A single set of eigenspectra are created for the entire SDSS
DR7. During development of this release, attempts were made to create
sets of eigenspectra for different plates, and for groups of fibres
which exhibit similar fluctuations. However, no substantial
improvement in the final result was found and therefore the simplest
(and original) method was adopted.

\subsection{Cleaning of object spectra}\label{sec:cleaning}

Once the eigensystem has been created from the sky spectra, the
second step of the procedure involves adding the components in linear
combinations to remove the systematic sky residuals from the spectra
of target objects, such as galaxies, quasars and stars.

The number of components to use in the reconstruction is not well defined.  The
use of too many components results in the artificial suppression of noise below
the Poisson limit, with the PCA acting as an (undesirable for most
science applications) high-frequency
filter.  The use of too few components means that the removal of sky residuals
is sub-optimal and, in some cases, the overall quality of the spectra can
decrease.  The reduction of the noise in the sky pixels below the noise in the
non-sky pixels is clearly unphysical, and we therefore estimate the number of
components to employ in the reconstruction of each spectrum by adopting the
non-sky pixels of that spectrum as a reference.  The number of components is
determined using the root-mean-square (RMS) of the (flux-continuum)/error array
for both the sky, and non-sky pixels.  All RMS-values are calculated after
subtraction of the object continuum, using a 71-pixel running median filter,
masking of potential line features as described below, and removal of bad
(error=0) pixels.  Despite these precautions, a robust estimator of the RMS is
still required, therefore the 90th percentile of the absolute
(flux-continuum)/error array is used as our ``RMS'' estimator.  This simplifies
slightly the procedure of WH05.  The error array used in this section has been
rescaled to empirically remove the rescaling applied during the SDSS pipeline
reduction at the position of the OH lines (see WH05 for more details).  All
three RMS values are included in the FITS-headers described in Section
\ref{sec:model}.

The reconstruction of a spectrum proceeds one component at a time,
with the RMS-ratio calculated between the sky-residual subtracted
spectrum and the non-sky pixels as reference.  The reconstruction is
stopped when the RMS-ratio reaches unity, i.e.  when the noise
weighted flux RMS is the same for the sky and non-sky pixels. The
scheme is largely self-calibrating.  For example, in spectra with high
SNR the systematic sky residuals typically contribute only marginally
to the sky-pixel noise, the RMS-ratio thus starts close to unity and
only a small number of components are necessary to achieve equality in
the RMS-ratio.

During projection of the eigenspectra onto the object spectra, bad pixels
(error=0) and pixels which may have been affected by a narrow emission or
absorption feature, not removed by the 71-pixel median filter, are masked
(Section \ref{sec:masks}).  However, OH emission line residuals are still
removed from these pixels via subtraction of the contiguous PCA components,
the amplitudes of which are determined using the rest of the wavelength range.  
The effective correction of pixels which include astrophysical features is a
major benefit of the PCA subtraction scheme compared to the application of
a high-pass filter or masking out pixels affected by OH residuals completely.

As in WH05 we have imposed an upper limit on the number of components
used of 300.  Any spectrum which reaches this upper limit (see header
keyword {\small NRECON} in the FITS files, see Table \ref{tab:header}) should be treated with
suspicion, as it is possible that the stopping criteria has failed due
to some real unmasked feature in the spectrum, such as strong
intervening absorption lines in quasars.  In a very small number of
spectra, the final reconstruction after 300 components has a sky pixel
RMS that is worse than prior to subtraction.  No correction is applied
to such spectra, which may be identified by inspection of the FITS
header keywords {\small SKYVAR1} and {\small SKYVAR2} (i.e. {\small
  SKYVAR}1={\small SKYVAR}2, see Table \ref{tab:header}).  The
reliability of the PCA reconstructions deteriorates significantly when
a substantial fraction of all the pixels are unavailable, therefore no
correction is applied to spectra where the combined fraction of bad
and masked pixels in the range $6700-9180\,$\AA \ exceeds 20 per cent.

\subsection{Samples}

In the inital public release (Version 1.0) we have made available
spectra for two science samples: galaxies and quasars.  Galaxies have
been selected as all objects in DR7 possessing spSpec FITS-files, with
header keyword {\small SPEC\_CLN}$=2$.  Quasars are selected from the
FITS header keyword {\small SPEC\_CLN}$=3$ or $4$.  Additionally, spectra of
additional objects not selected by the latter criterion, but included
in the DR7 quasar catalogue \citep{Schneider:2010p5407} have also been
treated as quasars.  Sky spectra ({\small SPEC\_CLN}$=5$) and spectra
without classification ({\small SPEC\_CLN}$=0$) have also been
processed and are included in the release.

\subsection{Masks}\label{sec:masks}

Narrow features, intrinsic to object spectra, that are not subtracted by the
median-filter derived ``continua'' (Section \ref{sec:cleaning}) can effect the determination of the sky- and
non-sky-pixel RMS values and bias the reconstruction of
the sky-residuals.  To mitigate both effects, narrow features, notably emission
and absorption lines, are masked during the projection of the eigenspectra and
calculation of the RMS values.  The masked pixels are still subject to the
sky-residual correction but their presence is not allowed to influence
the projection of the eigenspectra.  Details of the mask applied for each
spectrum are included in the third row of the FITS data files (see Table
\ref{tab:data}).

Individually tailored masks can be generated for each object spectrum, depending
on the specific science application.  However, for the majority of science
applications, the use of a liberally-defined mask results in only a minor
degredation in the quality of the sky-residual subtraction while minimising the
probability of undesirable bias, due to the influence of unmasked features on
the projections.

\begin{figure}
\includegraphics[scale=0.8]{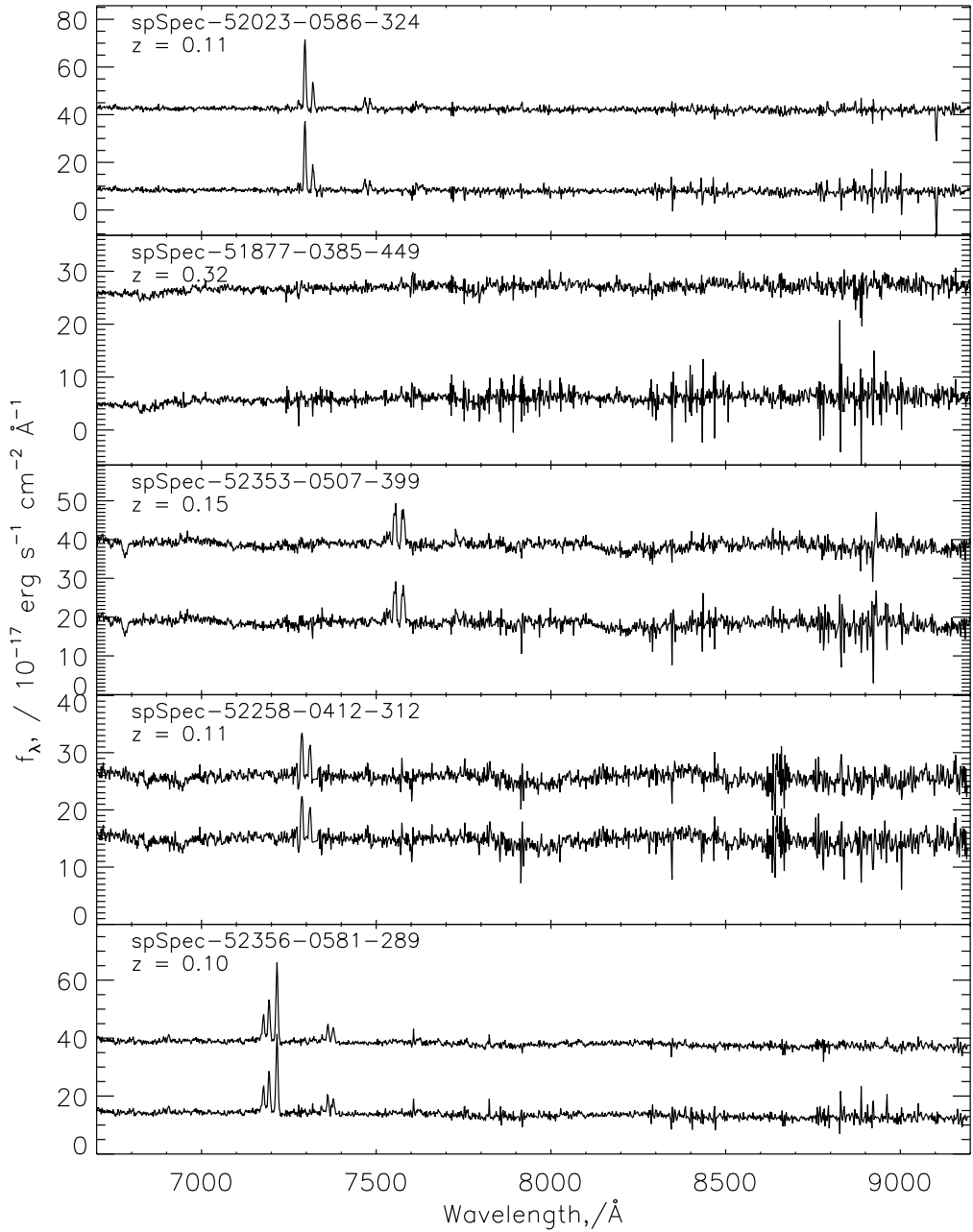}
\caption{Examples of sky residual subtraction applied to galaxy
  spectra. In each panel the lower spectrum is the raw SDSS data and
  the upper spectrum is after application of the sky-residual
  subtraction procedure. The upper sky-residual subtracted spectrum is
offset for clarity. }\label{fig:eggal}
\end{figure}

\begin{figure}
\includegraphics[scale=0.8]{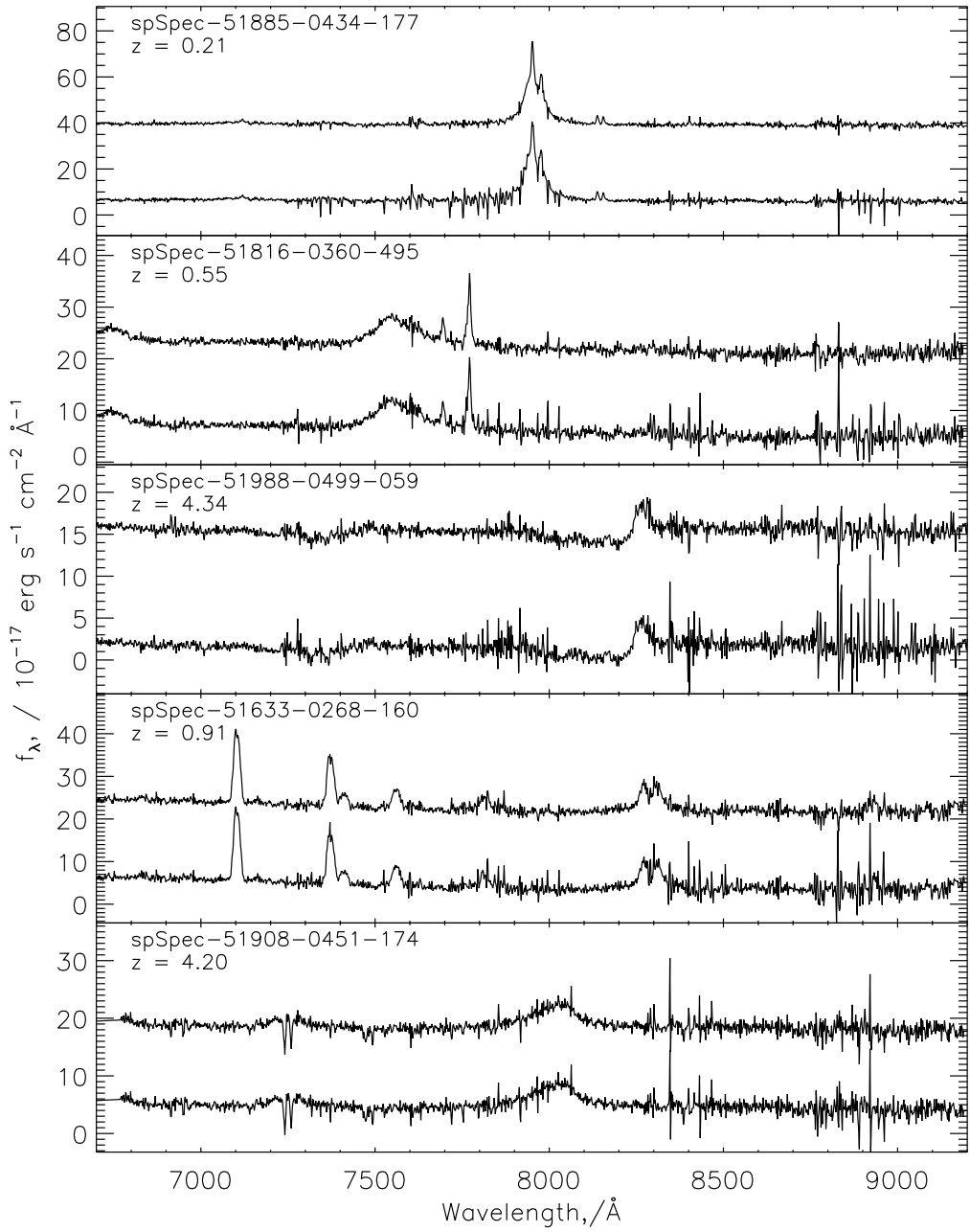}
\caption{Same as Figure \ref{fig:eggal}, but for quasars}\label{fig:egqso}
\end{figure}

\begin{figure}
\includegraphics[scale=0.7]{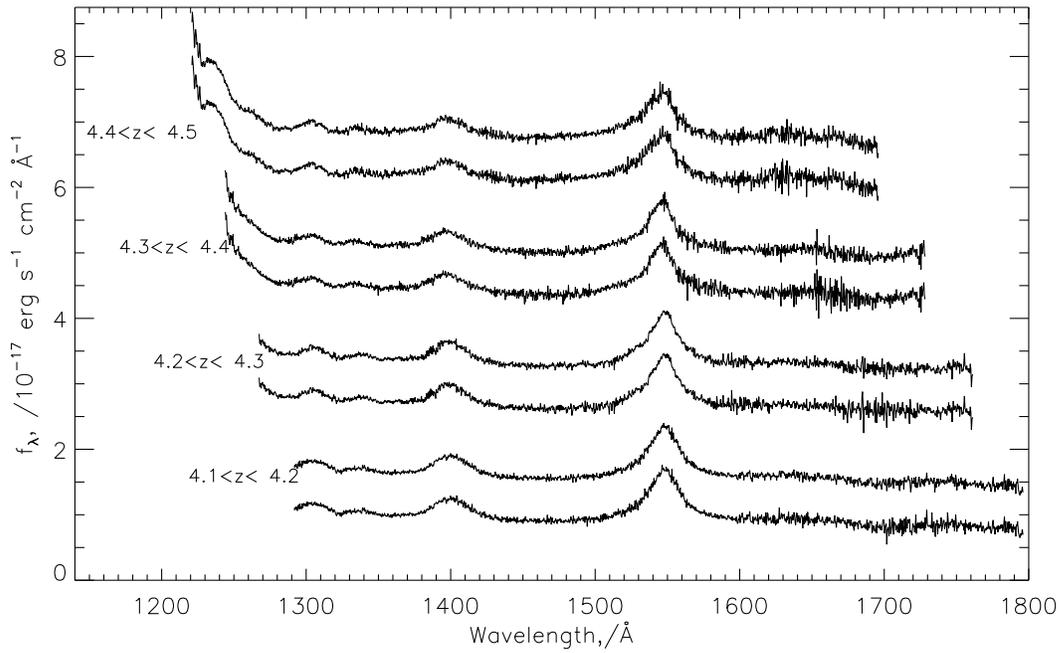}
\caption{Composite spectra of $z\sim4$ quasars. The observed frame
  spectra fall at wavelengths $>6700\AA$. For each pair of spectra,
  the lower composite is created directly from the SDSS spectra, and
  the upper composite from the same spectra after subtracting the sky
  residuals.}\label{fig:qsocomp}
\end{figure}

Thus, for galaxies, a single mask has been used which includes all the common
emission and absorption features found in galaxy spectra that are detectable at
the resolution and SNR of the SDSS spectra.  Details of the features, their
vacuum wavelengths and the rest-frame extent of the masked region are given in
Table \ref{tab:galwav}.  Redshifts, necessary to calculate the observed-frame
wavelengths of the masked regions, are taken from the SDSS FITS file primary
headers.  The same philosophy has been applied for the quasar sample although
redshifts are taken from the DR7 quasar catalogue of \citet{Schneider:2010p5407}
where available.  Vacuum wavelengths and the rest-frame extent of the masked
regions for quasars are given in Table \ref{tab:qsowav}\footnote{A region
centred on 4984\,\AA \ is masked to exclude artificial ``absorption'' resulting
from the continuum-subtracted spectra of objects possessing strong [O~{\sc
iii}]\,$\lambda\lambda$4960,5008 emission.}.  The advantage of using
the sky-residual subtracted spectra can be appreciated by inspection of Figs.
\ref{fig:eggal}, \ref{fig:egqso} and \ref{fig:qsocomp}.

It is important to recognise that the spectra resulting from the use
of the generic masking procedure described above are not suitable for
science investigations where additional narrow features are present at
a different redshift to the science object. Examples include quasar
absorption line catalog generation and the identification of spectra
consisting of multiple object components, such as galaxies with
gravitationally lensed background sources. The potential problems are
demonstrated via inspection strong MgII absorption systems at
$z_{abs}$=2.191 and $z_{abs}$=1.485 in the original and
sky-residual-subtracted spectra of spSpec-52282-0518-495 and
spSpec-52282-0328-472 respectively. In the former case, the absorption
system is virtually eliminated, while in the latter case, the absorber
doublet ratio alters dramatically.

\begin{table}
\centering
\caption{\label{tab:galwav} Galaxy Mask}
\vspace{0.2cm}
\begin{tabular}{ccl}\hline\hline
Central Wavelength & Size & Species\\
(\AA) & (\AA) & \\\hline
3728.30 & 20.0 &  O~{\sc ii}\\
3798.4 & 15.0 &  H$\theta$\\
3836.2 & 15.0 &  H$\eta$\\
3869.9 & 15.0 &  [Ne~{\sc iii}]\\
3889.00 & 15.0 & H$\zeta$ \\
3934.8 & 33.0 &  Ca~{\sc K}\\
3969.6 & 33.0 &  ca~{\sc H}\\
3971.19 & 15.0 & H$\epsilon$ \\
4102.9 & 20.0 &  H$\delta$\\
4305.6 & 30.0 &  G--band\\
4341.7 & 20.0 &  H$\gamma$\\
4364.44 & 15.0 & [O~{\sc iii}] \\
4472.5 & 15.0 &  He~{\sc i}\\
4862.68 & 30.0 &  H$\beta$\\
4960.30 & 15.0 &  [O~{\sc iii}]\\
5008.24 & 20.0 &  [O~{\sc iii}]\\
5176.7 & 30.0 &  Mg~{\sc b}\\
5200.0 & 10.0 &  [N~{\sc i}]\\
5272.0 & 20.0 &  \\
5877.4 & 15.0 &  He~{\sc i}\\
5891.58 & 20.0 & Na~{\sc i} \\
5897.56 & 20.0 & Na~{\sc i} \\
6302.3 & 15.0 &  [N~{\sc ii}]\\
6498.0 & 20.0 &  \\
6549.86 & 15.0 & [N~{\sc ii}] \\
6564.6 & 30.0 &  H$\alpha$\\
6585.27 & 15.0 & [N~{\sc ii}] \\
6680.4 & 15.0 &  \\
6718.3 & 20.0 &  [S~{\sc ii}]\\
6732.7 & 20.0 &  [S~{\sc ii}]\\
7066.4 & 15.0 &  He~{\sc i}\\
7138.4 & 15.0 &  [Ar~{\sc iii}]\\
7321.5 & 15.0 &  [O~{\sc ii}]\\
7331.6 & 15.0 &  [Ar~{sc iv}]\\
7753.5 & 10.0 &  [Ar~{\sc iii}]\\
8500.36 & 20.0 &  Ca~{\sc ii}\\
8544.38 & 25.0 &  Ca~{\sc ii}\\
8664.52 & 25.0 &  Ca~{\sc ii}\\
9018.0 & 15.0 &  \\
9072.0 & 15.0 &  [S~{\sc iii}]\\
\end{tabular}
\end{table}

\begin{table}
\centering
\caption{\label{tab:qsowav} Quasar Mask}
\vspace{0.2cm}
\begin{tabular}{ccl}\hline\hline
Central Wavelength & Size & Species\\
(\AA) & (\AA) & \\\hline
1216.0 & 40.0 & Ly$\alpha$ \\
1241.0 & 40.0 & N~{\sc v}\\
1305.0 & 40.0 & O~{\sc i}\\
1400.0 & 40.0 & Si~{\sc iv}+O~{\sc iv}\\
1549.0 & 40.0 & C~{\sc iv}\\
1909.0 & 40.0 & C~{\sc iii}]\\
2798.0 & 40.0 & Mg~{\sc ii}\\
3427.0 & 20.0 & [Ne~{\sc v}]\\
3729.0 & 20.0 & O~{\sc ii}\\
3870.0 & 20.0 & [Ne~{\sc iii}]\\
4103.0 & 40.0 & H$\gamma$ \\
4342.0 & 40.0 & H$\delta$ \\
4863.0 & 40.0 & H$\beta$ \\
4960.0 & 40.0 & [O~{\sc iii}]\\
4984.0 & 40.0 &  ---\\
5008.0 & 40.0 & [O~{\sc iii}]\\
5878.0 & 20.0 & He~{\sc i}\\
6302.0 & 20.0 & [O~{\sc i}]\\
6550.0 & 20.0 & [N~{\sc ii}]\\
6565.0 & 40.0 & H$\alpha$ \\
6585.0 & 20.0 & [N~{\sc ii}]\\
6725.0 & 40.0 & [S~{\sc ii}]\\
7139.0 & 20.0 & [Ar~{\sc iii}]\\
\end{tabular}
\end{table}

\section{Data model}\label{sec:model}

Conscious of the significant data volume of the SDSS Legacy Survey spectra we
have deliberately opted for a data model that minimises the duplication of
information contained in the original Legacy Survey FITS spSpec files.  Our new
files include a direct copy of the primary FITS header from the SDSS spSpec
file, with the addition of the FITS header keywords listed in Table
\ref{tab:header}.  Three {\small NAXIS1}-pixel floating-point data arrays, of
the same length as the original SDSS spectra, are included:  the sky-residual
subtracted spectrum flux array as the first row; the re-scaled error array as
the second row; the mask information as the third row (see Table
\ref{tab:data}).  Only those pixels where the sky-residual correction has been
applied have had their flux and error values altered, all other pixels remain
unchanged from the original SDSS spSpec file.

The sky-residual subtracted spectra have the same naming convention as the SDSS
spSpec files, i.e.  one file per spectrum, labeled by the mjd, plate and
fiberid of the spectrum.  The file names include the tag ``\_skysub'' to prevent
any confusion with the original SDSS spSpec files.  Individual files are stored
in directories according to their plate number.

The reduction of the number of data arrays per spectrum from five to three,
combined with the elimination of the extensive FITS-header units (containing
information about emission and absorption features detected in the spectrum)
results in a file size just under half that of the original
SDSS Legacy Survey spSpec files.

\begin{table}
\centering
\caption{\label{tab:header} FITS file primary header keywords}
\vspace{0.2cm}
\begin{tabular}{ll}\hline\hline
Keyword & Description \\\hline
{\small SKYVAR0}  & Non-sky pixel variance \\
{\small SKYVAR1} & Sky pixel variance before correction  error \\
{\small SKYVAR2} & Sky pixel variance after correction \\
{\small NRECON} & Number of components in reconstruction\\
\end{tabular}
\end{table}

\begin{table}
\centering
\caption{\label{tab:data} FITS file data content}
\vspace{0.2cm}
\begin{tabular}{clc}\hline\hline
Data row & Content & Equivalent SDSS data row\\\hline
1  & Object flux array & 1 \\
2 & Object error array & 3 \\
3 & Sky-correction mask & -- \\
\end{tabular}
\end{table}

\begin{table}
\centering
\caption{\label{tab:mask} FITS sky-correction mask values and their meanings}
\vspace{0.2cm}
\begin{tabular}{cl} \hline\hline
Mask value & Meaning \\\hline
0 & outside wavelength range $6700-9180\,$\AA \ for sky-correction  \\
1 & non-sky pixels: inside wavelength range but not subject to sky-correction \\
2 & sky pixels: inside wavelength range and subject to sky-correction \\
3 & masked non-sky pixels: pixel masked due to object feature or SDSS bad
pixel \\
4 & masked sky pixels: pixel masked due to object feature or SDSS bad
pixel \\
\end{tabular}
\end{table}

\subsection{Data Access}
The data files may be accessed through the Johns Hopkins University
SDSS server at the address

http://www.sdss.jhu.edu/skypca/spSpec 

Subsequent versions will be stored in the subdirectories v1/, v2/
etc. with the main directory given above pointing to the latest
version at all times.

The files are easily accessible via the ``wget'' command. Several
file lists have been prepared for the convenience of the user and can
be found in the main directory given above:
\begin{description}
\item[wget\_all.lis] To download all reduced spectra.
\item[wget\_dr7qsocat.lis] To download the spectra of the DR7 quasar catalog.
\item[wget\_specobj\_specclass2.lis] All spectra spectroscopically
  classified as galaxies ({\small SPECCLASS=2})
  in the CAS SPECOBJ view, except for 20 objects classified as
  galaxies in the CAS catalog, but not in their FITS file headers.  
\item[wget\_specobjall\_specclass2.lis] All spectra spectroscopically
  classified as galaxies ({\small SPECCLASS=2}) in the CAS SPECOBJALL
  table (i.e. includes duplicates), except for objects on plate 1631
  for which the spectra of the same mjd as in the catalog do not
  exist, and the 20 objects noted above.
\end{description}
After downloading the individual file list to your computer or
creating your own, a command such as:
\begin{verbatim}
wget -x -nH --cut-dir=2 -a logfile.txt -i wget_all.lis
\end{verbatim}
will download all the files in that file list to your
machine, creating the file logfile.txt to keep track of the progress. 
\vspace{1ex}

The DR7 QSOs may additionally be accessed from the DR7 Value Added
Catalogue web page:\\ http://www.sdss.org/dr7/products/value\_added/index.html\#quasars
\
\section{Conclusion}\label{sec:conclude}
We have released to the community a set of SDSS DR7 spectra with significantly
improved subtraction of OH sky lines.  The PCA-based technique employed takes
advantage of the correlation of OH lines as a function of wavelength and sky
subtraction close to the Poisson noise limit is achieved for many spectra.  Our
procedure is generally more effective than simply employing a high-pass filter
or simple feature-exclusion mask. Release Version 1.0 includes galaxies,
quasars, sky and ``unknown'' spectra. Future releases will incorporate improved
redshifts for the quasars and the masking of intervening absorption features in quasar spectra.

\section{Acknowledgements}
We would like to thank Alex Szalay and Tamas Budavari at Johns Hopkins
University for kindly organising the hosting of the data files for
public access, and Jan Vandenberg and Rich Ercolani for their time
setting up the web access. Gordon Richards enabled the spectra for the
SDSS DR7 quasar catalogue to be made available as part of the SDSS DR7
Legacy Release website `Added Value' data page.

\bibliographystyle{mn2e}
\bibliography{refs_all}

\begin{thebibliography}{}

\bibitem[\protect\citeauthoryear{Abazajian, Adelman-McCarthy, Ag{\"u}eros,
  Allam, Prieto, An, Anderson, Anderson \& et al.}{Abazajian
  et~al.}{2009}]{Abazajian:2009p5430}
Abazajian K.~N.,  Adelman-McCarthy J.~K.,  Ag{\"u}eros M.~A.,  Allam S.~S.,
  Prieto C.~A.,  An D.,  Anderson K. S.~J.,  Anderson S.~F.,    et al. 2009,
  The Astrophysical Journal Supplement, 182, 543

\bibitem[\protect\citeauthoryear{Budav{\'a}ri, Wild, Szalay, Dobos \&
  Yip}{Budav{\'a}ri et~al.}{2009}]{Budavari:2009p4530}
Budav{\'a}ri T.,  Wild V.,  Szalay A.~S.,  Dobos L.,    Yip C.-W.,  2009,
  Monthly Notices of the Royal Astronomical Society, 394, 1496

\bibitem[\protect\citeauthoryear{Schneider, Hall, Richards \& et al}{Schneider
  et~al.}{2007}]{2007AJ....134..102S}
Schneider D.~P.,  Hall P.~B.,  Richards G.~T.,    et al 2007, \aj, 134, 102

\bibitem[\protect\citeauthoryear{Schneider, Richards, Hall, Strauss, Anderson,
  Boroson, Ross, Shen \& et al.}{Schneider et~al.}{2010}]{Schneider:2010p5407}
Schneider D.~P.,  Richards G.~T.,  Hall P.~B.,  Strauss M.~A.,  Anderson S.~F.,
   Boroson T.~A.,  Ross N.~P.,  Shen Y.,    et al. 2010, The Astronomical
  Journal, 139, 2360

\bibitem[\protect\citeauthoryear{Wild \& Hewett}{Wild \&
  Hewett}{2005}]{2005MNRAS.358.1083W}
Wild V.,  Hewett P.~C.,  2005, \mnras, 358, 1083

\bibitem[\protect\citeauthoryear{York, Adelman, Anderson \& et al}{York
  et~al.}{2000}]{2000AJ....120.1579Y}
York D.~G.,  Adelman J.,  Anderson J.,    et al 2000, \aj, 120, 1579

\end{thebibliography}







\begin{appendix}
\section{Figures}
Here we include a repeat of some of the figures presented in WH05, for
easy reference, and to allow direct comparison between the results before and
after the re-reduction of the entire data set by the SDSS team. 

\begin{figure}[h]
\includegraphics[scale=0.4]{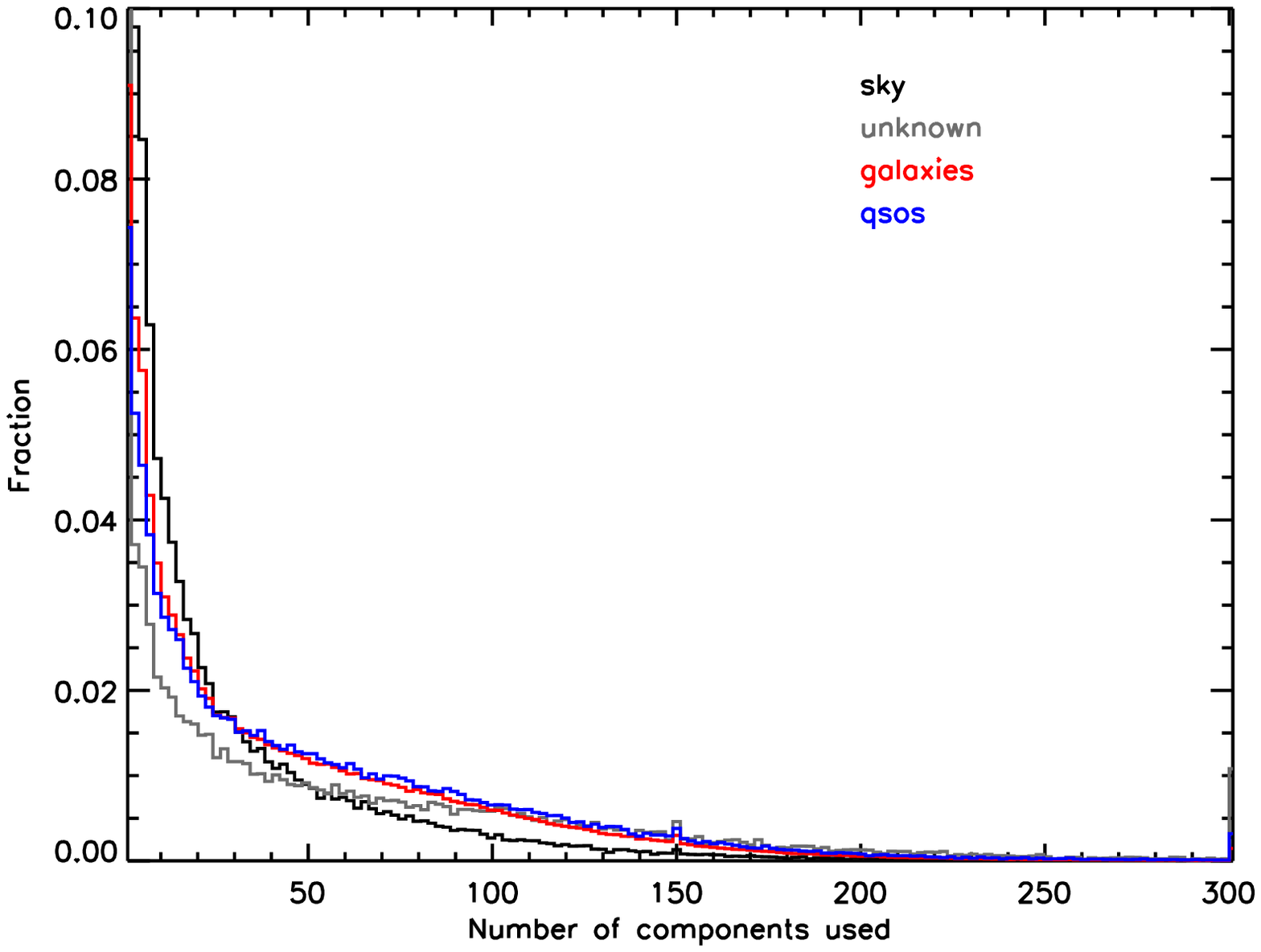}
\includegraphics[scale=0.4]{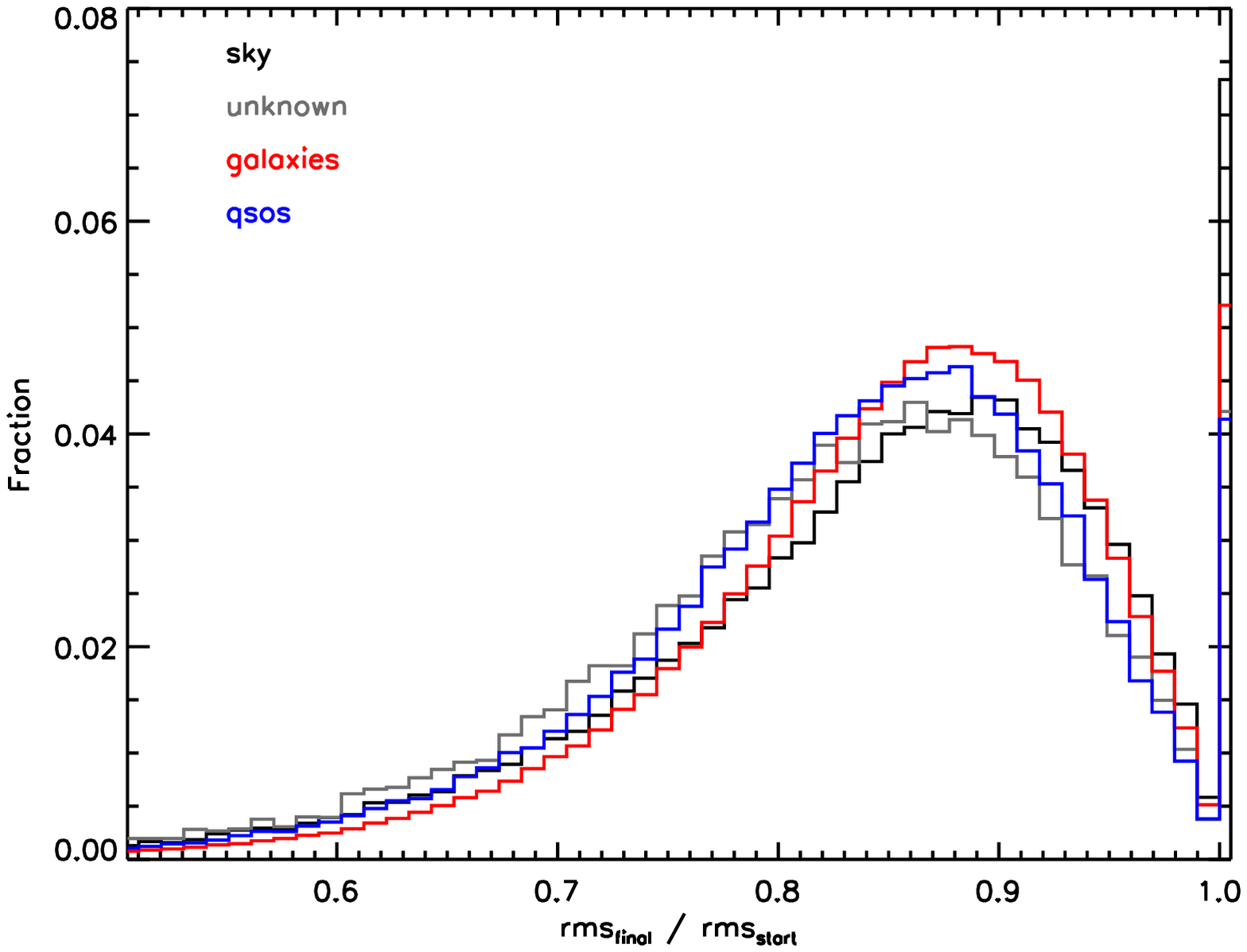}
\caption{{\it Left:} Histogram of the number of components used to
  reconstruct the sky signal in the sky, unknown, galaxy and quasar
  samples. The y-axis is truncated in order to show the majority of
  the objects. Twenty-three per cent of the unknown spectra require zero
  components, demonstrating the importance of employing appropriate 
  object-specific masks (Section \ref{sec:masks}). {\it Right:} 
  Histogram of the ratio between the root mean square noise of
  the pixels affected by OH lines before and after application of the
  sky-residual subtraction procedure. }\label{fig:nrecon}
\end{figure}

\end{appendix}

\end{document}